\documentclass[%
 reprint,
 amsmath,amssymb,
 aps,
]{revtex4-2}

\usepackage{graphicx}
\usepackage{dcolumn}
\usepackage{bm}
\graphicspath{{Figures/}}

\def\bs{\begin{subequations}}
\def\es{\end{subequations}}
\def\aa{\begin{align}}
\def\ab{\end{align}}
\def\ba{\begin{eqnarray}}
\def\ea{\end{eqnarray}}
\def\be{\begin{equation}}
\def\ee{\end{equation}}
\def\BS{\begin{split}}
\def\ES{\end{split}}

\def\ben{\begin{enumerate}}
\def\een{\end{enumerate}}
\def\bs{\bigskip}

\begin{document}

\preprint{APS/123-QED}

\date{\today}
\title{Enhanced diffusion in soft-walled channels with a periodically varying curvature}
\author{Thomas H. Gray$^1$, Claudio Castelnovo$^1$, Ee Hou Yong$^{2, *}$}
\affiliation{${}^1$ T.C.M. Group, Cavendish Laboratory, JJ Thomson Avenue, Cambridge, CB3 0HE, U.K.}
\affiliation{${}^2$Division of Physics and Applied Physics, School of Physical and Mathematical Sciences, Nanyang Technological University, Singapore 637371}
\email{eehou@ntu.edu.sg}

\begin{abstract}
In one-dimension, the diffusion of particles along a line is slowed by the addition of energy barriers. The same is true in two-dimensions, provided that the confining channel in which the particles move doesn't change shape. However, if the shape changes then this is no longer necessarily true; adding energy barriers can enhance the rate of diffusion, and even restore free diffusion. We explore these effects for a channel with a sinusoidally varying curvature.
\end{abstract}

\maketitle

\section{Introduction}
Transport processes in channels of varying profile have been studied for many years. With applications in disparate fields such as zeolites and porous solids \cite{RUTHVEN1995223, doi:10.1063/1.1480011}, biological membranes \cite{HILLE1978283, BEREZHKOVSKII2005L17, doi:10.1038/16426}, separating particles by their size \cite{doi:10.1021/cr960406n, PhysRevLett.108.020604,  doi:10.1063/1.4892615} and carbon nanotubes \cite{PhysRevLett.89.064503}, the importance of understanding these systems' behaviour is clear.

A common starting point is the Fick-Jacobs equation, an effective, one-dimensional equation for the evolution of the concentration of a solute along the centre-line of a multi-dimensional tube. Jacobs' treatment \cite{Jacobs1935}, which he attributed to Fick \cite{FickPaper}, was refined by Zwanzig \cite{doi:10.1021/j100189a004}, who produced a more general version. By assuming that the system is fully equilibrated in the direction normal to the length of the channel, Zwanzig reduced the multi-dimensional Smoluchowski equation to a one-dimensional form with a modified potential. The changing shape of the channel produces a logarithmic contribution to this potential, which leads to the description of the effect upon the motion in terms of `entropic' barriers \cite{PhysRevE.64.061106}.

However, it isn't safe to assume that the system equilibrates fully in the confining direction; a shape which varies too rapidly, for instance, can prevent equilibrium from being established \cite{doi:10.1063/5.0040071}. Zwanzig acknowledged the limitations of this approach and suggested how small deviations from equilibrium might be accounted for. This work, based around a spatially varying diffusion coefficient, has been built upon heavily \cite{doi:10.1063/1.1899150, PhysRevE.74.041203, PhysRevE.78.021103, BURADA200816}.

Introducing a linear bias along a periodic channel leads to motion described by a redefined effective diffusion coefficient and a non-linear mobility. The former peaks for some value of the applied force, before falling to a constant value; the latter increases monotonically towards a constant value \cite{PhysRevE.65.031104, PhysRevLett.96.130603, PhysRevE.75.051111}.

Although channels which are symmetrical about their axis feature prominently in this field, the more general case of a curved midline and varying width has also attracted attention. The motion can still be mapped onto one dimension, albeit with a modified expression for the spatially varying diffusion coefficient, which now reflects the variation in the midline of the channel \cite{PhysRevE.80.061142, doi:10.1063/1.3626215, doi:10.1063/1.4733394, DagdugMidline}. Motion in serpentine channels, where the midline is curved but the width is constant, has also been studied \cite{doi:10.1063/1.4917020, doi:10.1063/1.4940314}. 

Channels of curved midline and varying width can be created by using the same function to describe both walls, but then introducing a phase difference between the two. The current can be affected by this phase shift and a preferential direction of transport can emerge \cite{PhysRevE.87.062128, ALVAREZRAMIREZ2014319}.

Adding potential energy barriers to a channel of varying width can cause interesting effects because of the interaction between the energetic and entropic contributions to the potential. For instance, introducing a linear bias along the channel and tuning the phase difference between the periodic channel and the periodic barriers can induce a resonance-like behaviour in the non-linear mobility, and rectification can be observed \cite{BURADA2010514}. Another example involves a channel with cosine-shaped walls which connects two reservoirs of particles at different concentrations over one period. By introducing a cosine energy barrier along the channel, and tuning the phase relative to the walls, it is possible to produce transport from low to high concentration \cite{doi:10.1140/epjb/e2013-40862-y}. 

Here we study over-damped motion in a soft-walled channel whose profile varies periodically. By introducing a periodic potential along the channel it is possible to increase the rate of diffusion above that observed without the potential, in some cases up to free diffusion. The energy barriers enhance the motion.

\section{A one-dimensional model for the effective diffusion coefficient}
We will consider motion in a channel
\be
U\left(x,y\right) = U_x\left(x\right) + U_y\left(x,y\right),
\label{eq:U}
\ee
where $U_x$ is the potential energy contribution along the channel and $U_y$ describes how the profile of the channel varies as a function of the displacement along it. If the channel is periodic in $x$, then the long-time motion will be diffusive, and so described by an effective diffusion coefficient $D_\text{eff}$. We will use Zwanzig's derivation of the Fick-Jacobs equation to explore the system's behaviour. 

Zwanzig restricted his attention to the effect upon the diffusion coefficient of changes in the shape of the channel \cite{doi:10.1021/j100189a004}. Here we will retain the effect of energy barriers. Our starting point is the two-dimensional Smoluchowski equation for the probability density $p\left(x,y,t\right)$
\be
\begin{split}
\frac{\partial p}{\partial t} = \thinspace &D_x \frac{\partial}{\partial x} e^{-\beta U\left(x,y\right)} \frac{\partial}{\partial x} e^{\beta U\left(x,y\right)}\thinspace p \\
 &+ D_y \frac{\partial}{\partial y} e^{-\beta U\left(x,y\right)} \frac{\partial}{\partial y} e^{\beta U\left(x,y\right)}\thinspace p,
\end{split}
\label{eq:SE1}
\ee
where $D_x$ and $D_y$ are the free diffusion coefficients in the $x$ and $y$ directions, respectively. By inserting Eq.~(\ref{eq:U}) into Eq.~(\ref{eq:SE1}), integrating over the $y$-direction, and using the fact that $U_y$ is confining, we obtain
\be
\frac{\partial\rho}{\partial t} = \thinspace D_x\frac{\partial}{\partial x} e^{-\beta U_x} \int_{-\infty}^\infty \text{d}y \thinspace e^{-\beta U_y} \frac{\partial}{\partial x} e^{\beta U_x + \beta U_y} \thinspace p,
\label{eq:SE3}
\ee
where $\rho\left(x,t\right) = \int_{-\infty}^\infty \text{d}y \thinspace p\left(x,y,t\right)$ is the one-dimensional probability density. Let us assume that the distribution is always in equilibrium in the $y$-direction, i.e.
\be
p\left(x,y,t\right) \approx \rho\left(x,t\right) \frac{e^{-\beta U_y\left(x,y\right)}}{e^{-\beta A\left(x\right)}},
\label{eq:EqmApprox}
\ee
where $A\left(x\right)$ is defined through
\be
e^{-\beta A\left(x\right)} = \int_{-\infty}^\infty \text{d}y \thinspace e^{-\beta U_y\left(x,y\right)}.
\label{eq:ADef}
\ee
Inserting Eq.~(\ref{eq:EqmApprox}) into Eq.~(\ref{eq:SE3}) and carrying out the integration over $y$ produces the following partial differential equation for the one-dimensional density
\be
\frac{\partial\rho}{\partial t} = \thinspace D_x \frac{\partial}{\partial x} e^{-\beta U_x - \beta A} \frac{\partial}{\partial x} e^{\beta U_x + \beta A}\rho ,
\label{eq:SE5}
\ee
from which we can deduce the following expression for the one-dimensional effective potential $U^*$
\be
U^*\left(x\right) = U_x\left(x\right) - \frac{1}{\beta}\text{ln} \left(\int_{-\infty}^\infty \text{d}y \thinspace e^{-\beta U_y\left(x,y\right)}\right).
\label{eq:UstarDef}
\ee


Before we restrict our attention to a particular channel it is worth remarking upon an implication of Eq.~(\ref{eq:UstarDef}). Variations in the shape of the channel impede motion, a feature accounted for by the second term in the expression for the effective potential. However, Eq.~(\ref{eq:UstarDef}) implies that this retarding effect can be eradicated by introducing a potential in the $x$-direction: by setting $U_x = \frac{1}{\beta}\text{ln}\int_{-\infty}^\infty \text{d}y \thinspace e^{-\beta U_y\left(x,y\right)}$ the effective potential is zero and free diffusion is predicted. This is a point to which we will return.

We will now focus on the potential energy landscape
\be
\begin{split}
U_x\left(x\right) &= \frac{Q}{2}\left[1+\text{cos}\left(\frac{2\pi}{L}\left(x-\Delta x\right) \right)\right], \\
U_y\left(x,y\right) &= \frac{1}{2}\left[\alpha_0 + \alpha_1\text{cos}\left(\frac{2\pi x}{L}\right)\right]y^2,
\end{split}
\label{eq:UyDef}
\ee
where $Q\geq0$, and $\alpha_0>\alpha_1$ to make the channel confining, and study the effects of $Q$ and $\Delta x$ on the motion.

With the expression for the one-dimensional effective potential in Eq.~(\ref{eq:UstarDef}), we can derive the effective diffusion coefficient by considering the mean first-passage time from the potential energy maximum at $x=\Delta x$ to either of the maxima at $x=\Delta x \pm NL$. This is given by
\be
\begin{split}
\tau_N &= \frac{\mathcal{P}_R}{D_x} \int_{\Delta x}^{\Delta x +NL} \text{d}y \thinspace e^{\beta U^*\left(y\right)} \int_{\Delta x-NL}^y \text{d}z \thinspace e^{-\beta U^*\left(z\right)} \\
&- \frac{\mathcal{P}_L}{D_x} \int_{\Delta x-NL}^{\Delta x} \text{d}y \thinspace e^{\beta U^*\left(y\right)} \int_{\Delta x-NL}^y \text{d}z \thinspace e^{-\beta U^*\left(z\right)},
\end{split}
\label{eq:Tau1}
\ee
where $\mathcal{P}_L$ and $\mathcal{P}_R$ are the probabilities that the particle exits the region $[\Delta x-NL, \Delta x+NL]$ to the left and right, respectively \cite{GardinerHandbook}. The symmetry of the energy landscape means that $\mathcal{P}_L = \mathcal{P}_R = 1/2$, and Eq.~(\ref{eq:Tau1}) simplifies to
\be
\tau_N = \frac{N^2}{2D_x} \int_{\Delta x}^{\Delta x+L} \text{d}y \thinspace e^{\beta U^*\left(y\right)} \int_{\Delta x-L}^{\Delta x} \text{d}z \thinspace e^{-\beta U^*\left(z\right)},
\label{eq:Tau2}
\ee
where we have used the periodicity of the potential to recast each integral over one period.

After evaluating Eq.~(\ref{eq:UstarDef}) for the potential defined in Eq.~(\ref{eq:UyDef}), inserting the result into Eq.~(\ref{eq:Tau2}), and changing variables to $\theta = 2\pi x/L$, we find
\be
\tau_N = \frac{\left(NL\right)^2}{8\pi^2 D_x} \mathcal{I}_- \mathcal{I}_+,
\label{eq:Tau3}
\ee
where the quantities $\mathcal{I}_\pm$ are given by
\be
\mathcal{I}_\pm = \int_\phi^{2\pi+\phi}\text{d}\theta \thinspace e^{\pm\frac{\beta Q}{2}\text{cos}\left(\theta-\phi\right)} \left[\alpha_0 + \alpha_1 \text{cos}\thinspace\theta\right]^{\pm\frac{1}{2}},
\label{eq:Idef}
\ee
and $\phi=2\pi\Delta x/L$ is the phase difference. Finally, we obtain the effective diffusion coefficient 
\be
D_\text{eff} = \lim_{N\rightarrow\infty} \frac{\left(NL\right)^2}{2\tau_N} = \frac{4\pi^2 D_x}{\mathcal{I}_- \mathcal{I}_+}.
\label{eq:Deff1}
\ee

The derivative of $D_\text{eff}$ with respect to $\beta Q$ can be an informative quantity. For a one-dimensional system it is at most zero, and it is negative for $\beta Q>0$; increasing the height of the energy barriers reduces the size of the effective diffusion coefficient. Our quasi-one-dimensional system displays a more complicated behaviour:
\be
\frac{\partial D_\text{eff}}{\partial\beta Q}\bigg|_{\beta Q=0} = -\frac{D_0 \thinspace \mathcal{C}\left(\alpha_0, \alpha_1\right)}{2} \text{cos}\thinspace \phi,
\label{eq:PartialDeff4}
\ee
where $D_0=D_\text{eff}\left(\beta Q=0\right)$, and $\mathcal{C}\left(\alpha_0, \alpha_1\right)$ is a positive constant given by
\be
\begin{split}
\mathcal{C}\left(\alpha_0,\alpha_1\right) = &\frac{\int_0^{2\pi} \text{d}\theta \thinspace \text{cos}\thinspace \theta\sqrt{\alpha_0+\alpha_1\text{cos}\thinspace \theta}}{\int_0^{2\pi} \text{d}\theta \thinspace \sqrt{\alpha_0+\alpha_1\text{cos}\thinspace \theta}} \\
&- \frac{\int_0^{2\pi} \text{d}\theta \thinspace \text{cos}\thinspace \theta/\sqrt{\alpha_0+\alpha_1\text{cos}\thinspace \theta}}{\int_0^{2\pi} \text{d}\theta \thinspace /\sqrt{\alpha_0+\alpha_1\text{cos}\thinspace \theta}}.
\end{split}
\label{eq:IntegralDef}
\ee
A central result of our work is that Eq.~(\ref{eq:PartialDeff4}) can be positive; adding energy barriers can enhance the rate of diffusion along the channel. 

\section{Brownian dynamics simulations}
We used numerical simulations to study the two aspects of this work: the effect of the phase difference $\phi$ upon the behaviour of the effective diffusion coefficient, and the exact cancellation of energetic and entropic barriers to motion. Unless stated otherwise, simulations were performed with $10^5$ particles, a time step $\delta t = 10^{-4}$ units, and unit values of the thermal energy $k_\text{B}T$ and damping coefficients $\gamma_x$ and $\gamma_y$.

Eq.~(\ref{eq:PartialDeff4}) predicts that the gradient of the effective diffusion coefficient at $\beta Q=0$ is proportional to $\text{cos}\thinspace\phi$. We performed simulations for $\phi=0$ and $\phi=\pi$ to investigate the extremal cases. We will start with the former, because the behaviour is familiar.

\begin{figure}[h]
\includegraphics[width=1\columnwidth]{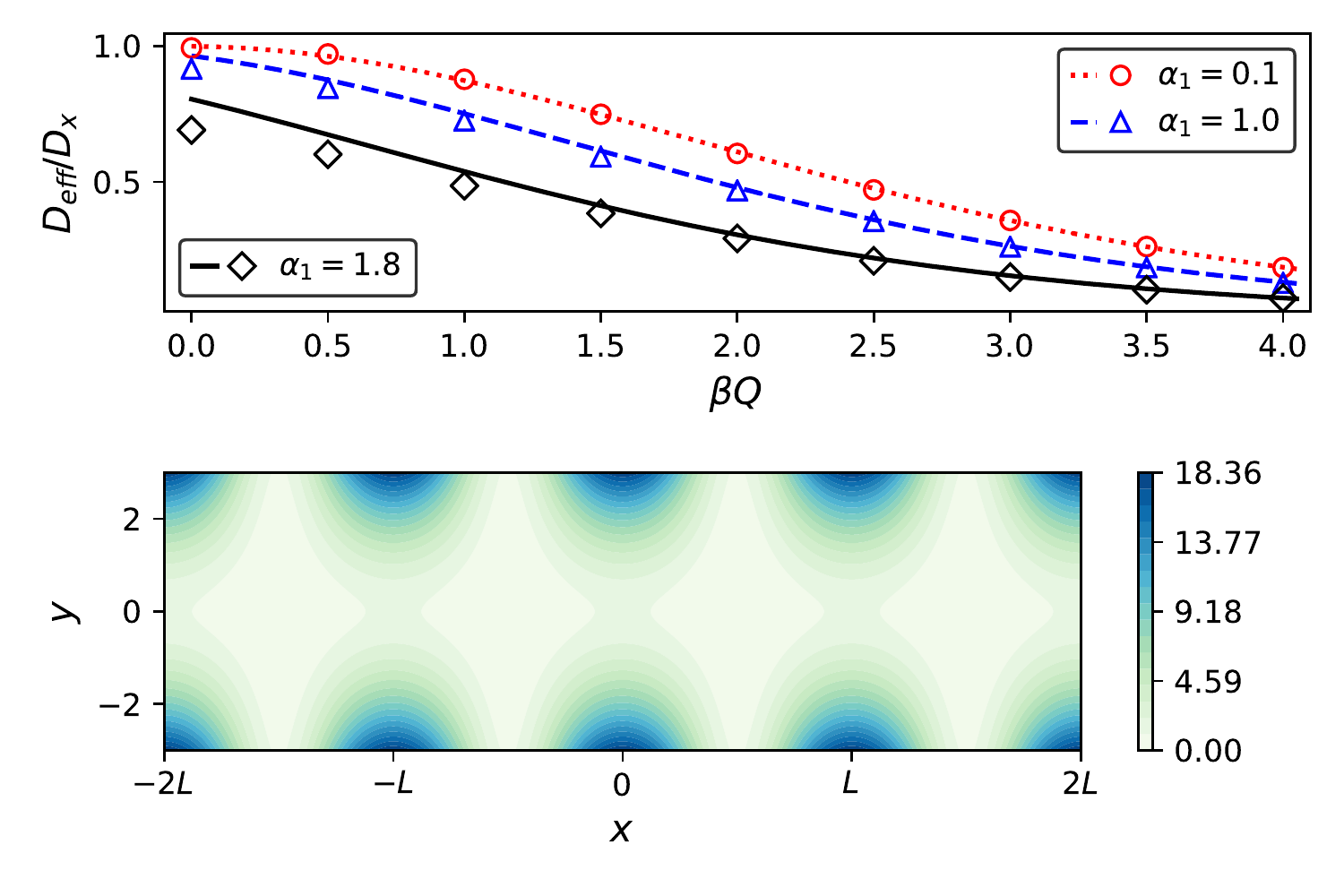}
\caption{The effective diffusion coefficient $D_\text{eff}$ is plotted as a function of the cosine barrier height for three values of $\alpha_1$. The phase difference is $\phi = 0$, and $\alpha_0=2$. The lines represent the theory -- Eq.~(\ref{eq:Deff1}) -- whilst the symbols represent the results of the numerical simulations. The contour plot is for $\beta Q=1.25525, \alpha_1=1.8$.}
\label{fig:DeffDeltaX05}
\end{figure}

We expect a negative gradient at $\beta Q=0$, and hence a monotonically decreasing effective diffusion coefficient. Fig.~(\ref{fig:DeffDeltaX05}) confirms these expectations and reveals good agreement between theory and simulations over a range of amplitudes. This is because the potential energy minima -- around which the particles spend the bulk of their time -- coincide with the points of minimum curvature. The distribution can get closer to its equilibrium shape in the regions of the channel where it would otherwise struggle most to do so. Agreement improves with increasing amplitude because more time is spent around the minima. Finally, agreement is better for smaller values of $\alpha_1$ because the variations in curvature are smaller.


\begin{figure}[h]
\includegraphics[width=1\columnwidth]{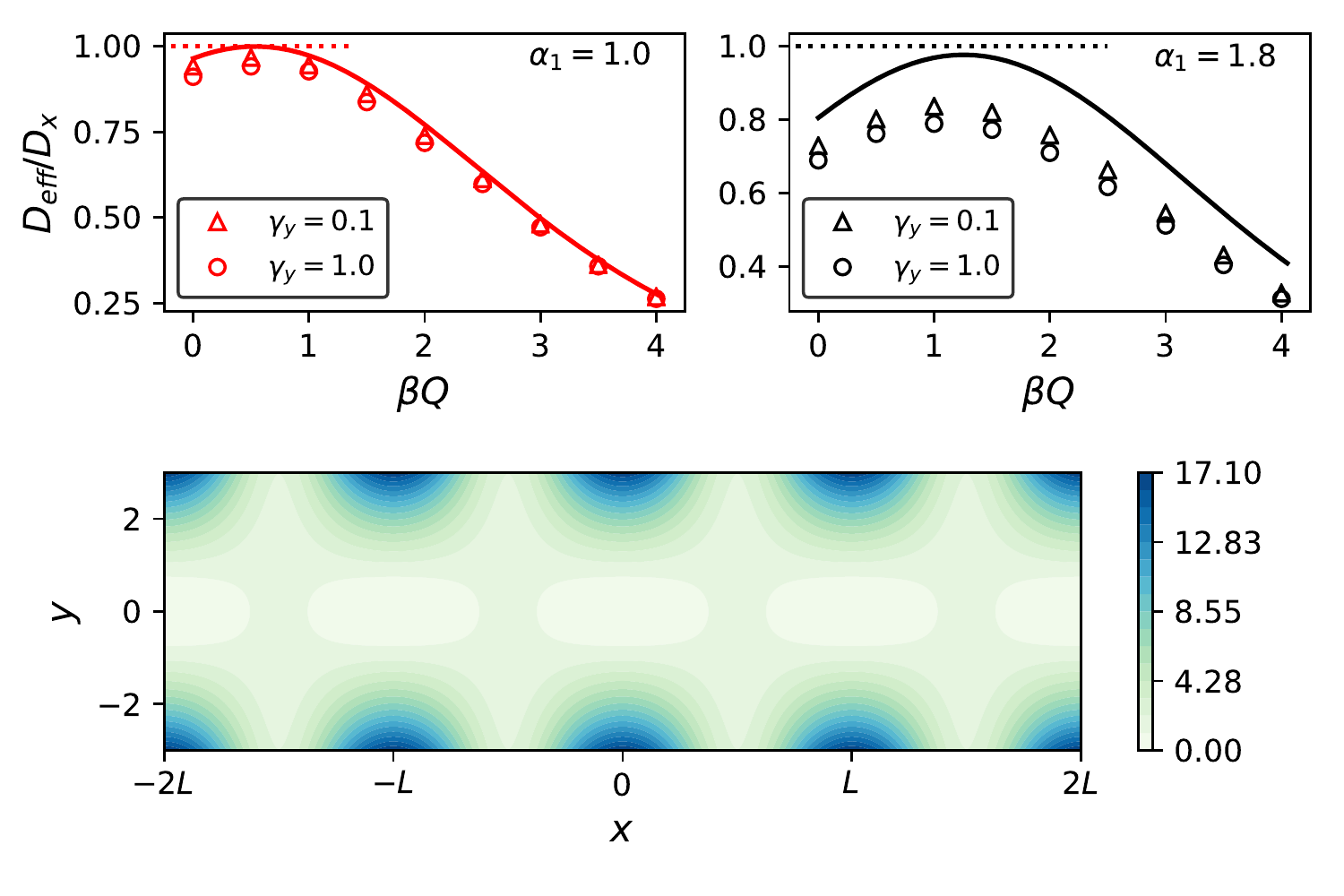}
\caption{The effective diffusion coefficient $D_\text{eff}$ is plotted as a function of the cosine barrier height for two values of $\alpha_1$, and two values of $\gamma_y$ in each case. The phase difference is $\phi = \pi$ , and $\alpha_0=2$. The lines represent the theory -- Eq.~(\ref{eq:Deff1}) -- whilst the symbols represent the results of the numerical simulations. The contour plot is for $\beta Q=1.25525, \alpha_1=1.8$.}
\label{fig:DeffDeltaX00}
\end{figure}

Let us now turn to the case $\phi=\pi$. Eq.~(\ref{eq:PartialDeff4}) predicts that the effective diffusion coefficient initially grows with the amplitude of the potential energy barriers. Fig.~(\ref{fig:DeffDeltaX00}) confirms this and reveals qualitative agreement between theory and simulations. However, quantitative agreement is not as good as in the previous case. This is because the points of minimum curvature coincide with the potential energy maxima. These are unstable points and particles pass through them quickly, leaving little chance for the ensemble to equilibrate. When $\alpha_1=1$ we see good quantitative agreement with the theory for values of $\beta Q>2$. In contrast, for $\alpha_1=1.8$ there is a lack of good agreement even for $\beta Q=4$. This is because the size of the entropic barriers to motion increases with the variation in the curvature of the channel, which is controlled by $\alpha_1$. For smaller values of $\alpha_1$, the rate of diffusion along the channel becomes determined by the height of the potential energy barriers at smaller values of the barrier height. Once in this regime, equilibration is less important for close agreement with the theory. As expected, decreasing $\gamma_y$ improves agreement with the theory.


\begin{figure}[h]
\includegraphics[width=1\columnwidth]{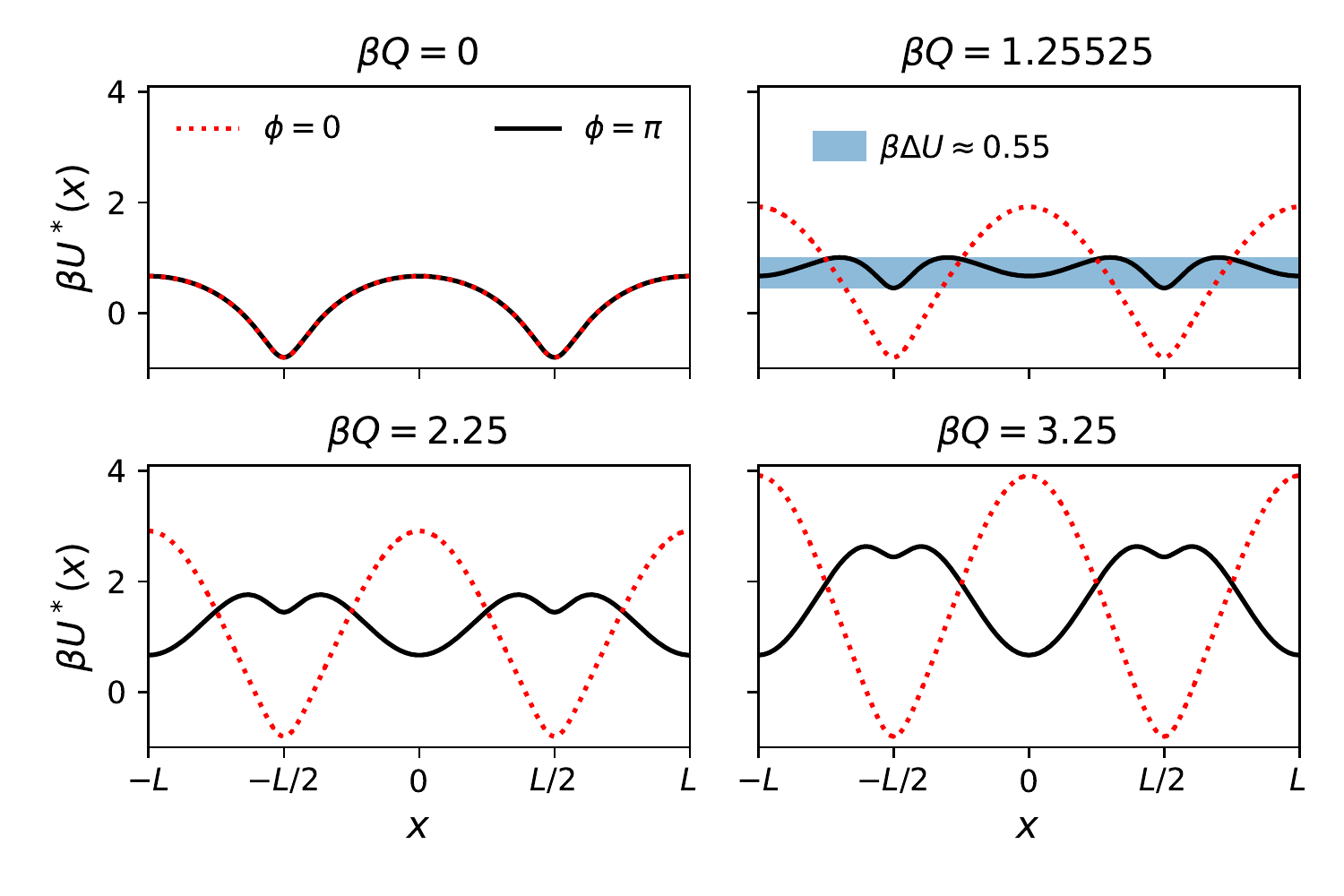}
\caption{The effective potential -- Eq.~(\ref{eq:UstarDef}) -- is plotted for the landscape described in Eq.~(\ref{eq:UyDef}) for a series of values of the cosine barrier height for two values of the phase $\phi$ in each case. $\alpha_1=1.8, \alpha_0=2$.}
\label{fig:Landscapes}
\end{figure}

Fig.~(\ref{fig:Landscapes}) provides insight into the origin of the behaviour of the effective diffusion coefficient: increasing the amplitude of the cosine potential does not necessarily increase the barrier to motion in the effective potential. The energetic and entropic contributions can interact with one another so as to \textit{decrease} the barrier to motion, as can be seen by comparing the panels for $\beta Q=0$ and $\beta Q=1.25525$, chosen because it is a good approximation to the amplitude which minimises the barrier.


The contour plots in Fig.~(\ref{fig:DeffDeltaX05}) and Fig.~(\ref{fig:DeffDeltaX00}) further our understanding of this effect. Fig.~(\ref{fig:DeffDeltaX05}) reveals that introducing the cosine potential creates near-flat regions which extend away from the centre of the channel. By contrast, the near-flat regions in Fig.~(\ref{fig:DeffDeltaX00}) extend much further along the line of the channel than away from it. The former will inhibit motion along the channel by enabling particles to move significant distances in unproductive directions. The latter comes close to providing a continuous near-flat region along the line of the channel, which is combined with steeper barriers to motion away from it. The region either side of the centre-line is flatter in Fig.~(\ref{fig:DeffDeltaX00}) than in Fig.~(\ref{fig:DeffDeltaX05}), and the saddle points are broader, which is beneficial for transport; it is easier for particles to move from a tighter minimum into a broader saddle than vice versa.

Let us conclude this section by returning to a point made after the introduction of the effective potential in Eq.~(\ref{eq:UstarDef}). The potential energy landscape
\be
U\left(x,y\right) = -\frac{1}{2\beta}\text{ln}\left(\frac{\beta \alpha\left(x\right)}{2\pi}\right) + \frac{1}{2}\alpha\left(x\right)y^2
\label{eq:UstarFreeDiff}
\ee
has been constructed by adding to the $U_y$ term describing the shape of the channel a series of potential energy barriers in the $x$-direction such that the effective potential is exactly zero. This predicts free diffusion.

\begin{figure}[h]
\includegraphics[width=1\columnwidth]{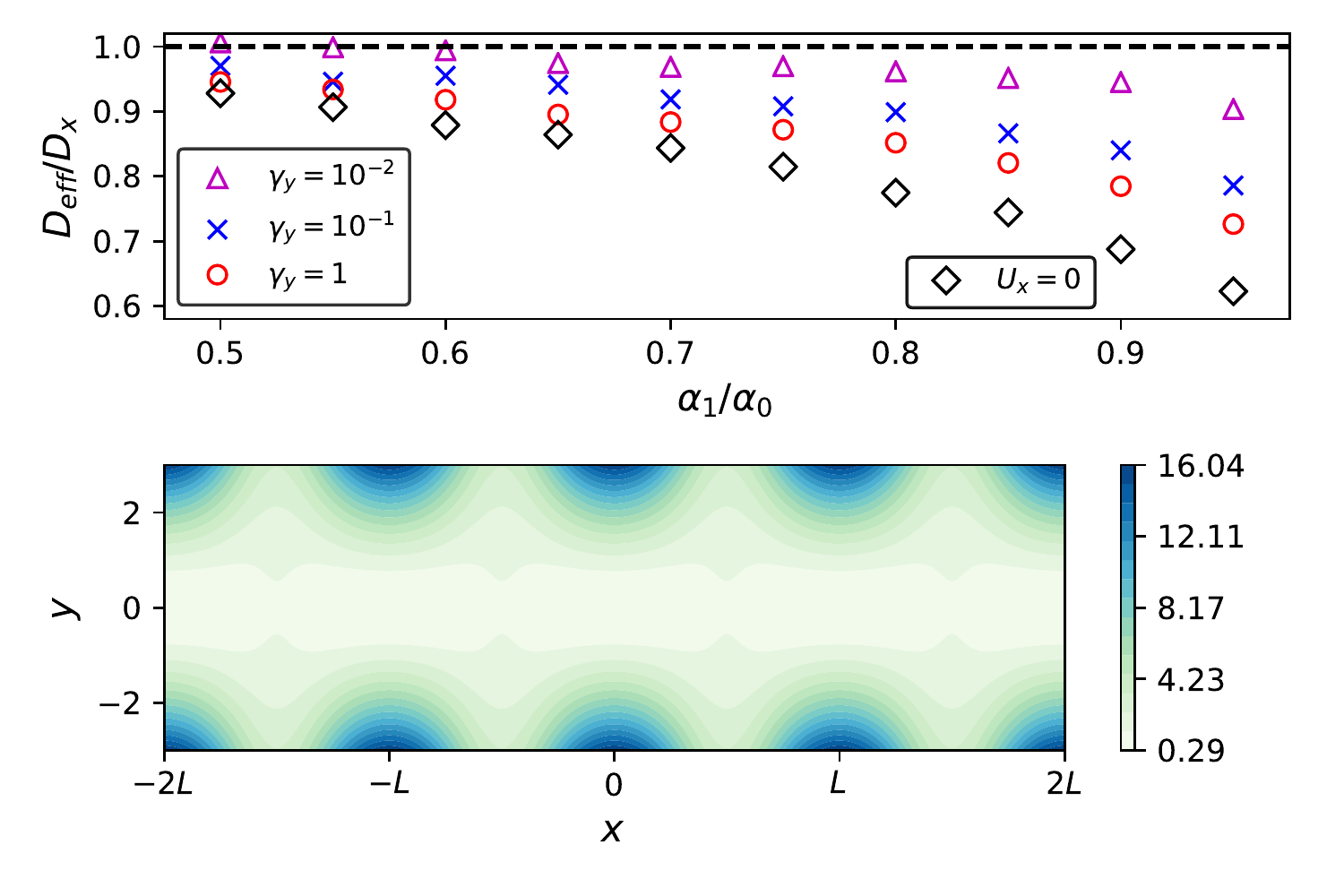}
\caption{The effective diffusion coefficient $D_\text{eff}$ for motion in the potential described in Eq.~(\ref{eq:UstarFreeDiff}) is plotted as a function of $\alpha_1/\alpha_0$ for a series of values of $\gamma_y$ ($\alpha_0=2$ throughout). The effective diffusion coefficient in the absence of energy barriers is also plotted $\left(U_x=0\right)$. The contour plot $\left(\alpha_1=1.5, \alpha_0=2\right)$ reveals that the channel has a near-flat central section along $x$.}
\label{fig:DeffFree}
\end{figure}

Motion was simulated for a range of values of $\alpha_1$ for $\alpha_0=2$. In each case, the effect of decreasing $\gamma_y$ from one to $10^{-2}$ upon the effective diffusion coefficient was studied. Motion was also simulated in the absence of the energy barriers. The results are shown in Fig.~(\ref{fig:DeffFree}).

For all values of $\alpha_1$ the effective diffusion coefficient is larger in the presence of energy barriers than in their absence. Again, as expected, decreasing $\gamma_y$ improves agreement with the theory.


\section{Conclusions}
We used the Fick-Jacobs equation to study the behaviour of particles diffusing along channels with a periodically varying profile and potential energy barriers along their length. Treating the variations in the shape of the channel as entropic barriers to motion reduces the problem to diffusion in an (approximate) one-dimensional potential.


For the cosine-based potential studied here, the position of the potential energy minima relative to the points of minimum curvature determines how the effective diffusion coefficient responds to increasing the height of the energy barriers. If the two coincide, then a monotonic decrease is observed, and there is good quantitative agreement with the theory. If the two are perfectly out-of-phase, so that the energy minima coincide with the regions of maximum curvature, then the effective diffusion coefficient initially increases above its zero-amplitude value, resulting in enhanced diffusion. Good quantitative agreement is observed only when the energy barriers dwarf the entropic barriers. Before this point, lack of equilibration in the confining direction precludes good agreement. 

For a given channel it is possible to construct a series of energy barriers which cancel out the entropic barriers; free diffusion is then predicted. Numerical simulations confirm that adding these barriers increases the rate of diffusion, and decreasing the damping coefficient in the confining direction leads ever-closer to free diffusion.

\section{Acknowledgements}
T.H.G acknowledges support from the EPSRC and E.H.Y. acknowledges support from Nanyang Technological University, Singapore, under its Start Up Grant Scheme (04INS000175C230).

\bibliography{Rough}

\end{document}


\date{\today}
\title{Supplemental material}
\author{Thomas H. Gray$^1$, Claudio Castelnovo$^1$, Ee Hou Yong$^{2, *}$}
\affiliation{${}^1$ T.C.M. Group, Cavendish Laboratory, JJ Thomson Avenue, Cambridge, CB3 0HE, U.K.}
\affiliation{${}^2$Division of Physics and Applied Physics, School of Physical and Mathematical Sciences, Nanyang Technological University, Singapore 637371}
\email{eehou@ntu.edu.sg}

\maketitle

\section{Connection to hard-walled potential}
In this work we chose to restrict our focus to soft-walled channels where the confining potential in the $y$-direction permits particles to explore out to $y=\pm\infty$. However, we don't believe that this places severe restrictions on the generality of the work. Indeed, the hard-walled potential, which features more commonly in the literature, can be viewed as a limiting form of the soft-walled potential we studied.

Let us begin by considering the expression for the effective potential $U^*\left(x\right)$ used in the manuscript
\be
U^*\left(x\right) = U_x\left(x\right) - \frac{1}{\beta} \text{ln}\left(\int_{-\infty}^\infty \text{d}y \thinspace e^{-\beta U_y\left(x,y\right)}\right).
\label{eq:UstarDef}
\ee
The limits on the integral reflect the fact that the potential is defined out to $y=\pm\infty$. However, if we consider a hard-walled potential which is flat between $y=\pm w\left(x\right)/2$ and infinite otherwise, then the expression for the effective potential becomes
\be
\begin{split}
U^*\left(x\right) &= U_x\left(x\right) - \frac{1}{\beta} \text{ln}\left(\int_{-w\left(x\right)/2}^{w\left(x\right)/2} \text{d}y\right), \\
&= U_x\left(x\right) - \frac{1}{\beta}\text{ln}\left[w\left(x\right)\right].
\end{split}
\label{eq:UstarDef2}
\ee

Having derived the hard-wall expression for the effective potential, let us draw the link between the soft- and hard-walled potentials. We chose to work with the potential described by
\be
U_y\left(x,y\right) = \frac{1}{2}\alpha\left(x\right)y^2.
\label{eq:UyDef}
\ee
Suppose that we modify this expression to
\be
U_y\left(x,y\right) = \frac{1}{2}\left[\alpha\left(x\right) y^2\right]^m,
\label{eq:UyDef2}
\ee
where $m$ is a positive integer, and then rearrange it into the following form
\be
U_y\left(x, y\right) = \frac{1}{2}\left[\left(\frac{y}{w\left(x\right)/2}\right)^2\right]^m,
\label{eq:UyMod}
\ee
where we have defined $w/2 = \alpha^{-1/2}$.

Inserting Eq.~(\ref{eq:UyMod}) into the integral in Eq.~(\ref{eq:UstarDef}) we find
\be
\begin{split}
\int_{-\infty}^\infty \text{d}y \thinspace e^{-\beta U_y} &= \int_{-\infty}^\infty \text{d}y \thinspace e^{-\frac{\beta}{2}\left(\frac{y}{w/2}\right)^{2m}}, \\
&= \frac{w}{2} \int_{-\infty}^\infty \text{d}z \thinspace e^{-\beta z^{2m}/2},
\end{split}
\label{eq:Int1}
\ee
where in moving from the first line to the second we have changed variables to $y=z/(w/2)$. It can be demonstrated that the limiting value of the integral in Eq.~(\ref{eq:Int1}) as $m\rightarrow\infty$ is two. Using this in Eq.~(\ref{eq:UstarDef}), one obtains the same effective potential as in Eq.~(\ref{eq:UstarDef2}).



By drawing the link between hard- and soft-walled potentials we hope that our results derived for the case of a soft-walled potential are interpreted more broadly.

\bibliography{Rough}
\bibliographystyle{apsrev4-2.bst}